\documentclass{kapproc}
\usepackage{t1enc}
\usepackage{procps} 
\usepackage[dvips]{graphicx}
\upperandlowercase
\kluwerbib
\input{psfig}
\def\deg{\mbox{$^{\mbox{o}}$\ }}
\def\30{30\,Dor}
\def\h2{{\sc Hii}}

\begin{document}

\articletitle{30 Doradus -- \\
              a Template for ``Real Starbursts'' ?}


\author{Bernhard R. Brandl}
\affil{Leiden Observatory, P.O. Box 9513, 2300 RA Leiden, The Netherlands}
\email{brandl@strw.leidenuniv.nl}

\chaptitlerunninghead{30\,Doradus -- a starburst template}

\anxx{Brandl\, Bernhard R.}

\begin{abstract}
\30 is the closest massive star forming region and the best studied
template of a starburst.  In this conference paper we first summarize
the properties of \30 and its stellar core R136.  We discuss the
effects of insufficient spatial resolution and cluster density
profiles on dynamical mass estimates of super star clusters, and show
that their masses can be easily overestimated by a factor of ten or
more.  From a very simple model, with R136-like clusters as
representative building blocks, we estimate typical luminosities of
the order $10^{11} L_\odot$ for starburst galaxies.
\end{abstract}



\section{Overview}

At a distance of only about 53~kpc \30 is the closest massive star
forming region.  To date it has been studied and described in over
3000 papers.  In this conference paper we discuss the relevance of \30
as a local template for more luminous starburst systems.  The
characteristic dimensions for structures related to \30 are given in
Tab.~\ref{sizes}.

\small
\begin{table}[ht]
\caption{Characteristic dimensions, from \cite{wal91}\label{sizes}}
\begin{tabular*}{7.5cm}{llcr}
\sphline
\it Name &\it Class &\it angular \O &\it linear \O \cr
\sphline
LMC & galaxy & 5\deg & 5000\,pc \cr
\30 region & complex & 1\deg & 1000\,pc \cr
\30 nebula & \h2 region & $15'$ & 200\,pc \cr
NGC\,2070 & stellar cluster & $3'$ & 40\,pc \cr
R136 & stellar core & $10''$ & 2.5\,pc \cr
\sphline
\end{tabular*}
\end{table}
\normalsize

\begin{figure}[ht]
\centerline{\hbox{
\psfig{figure=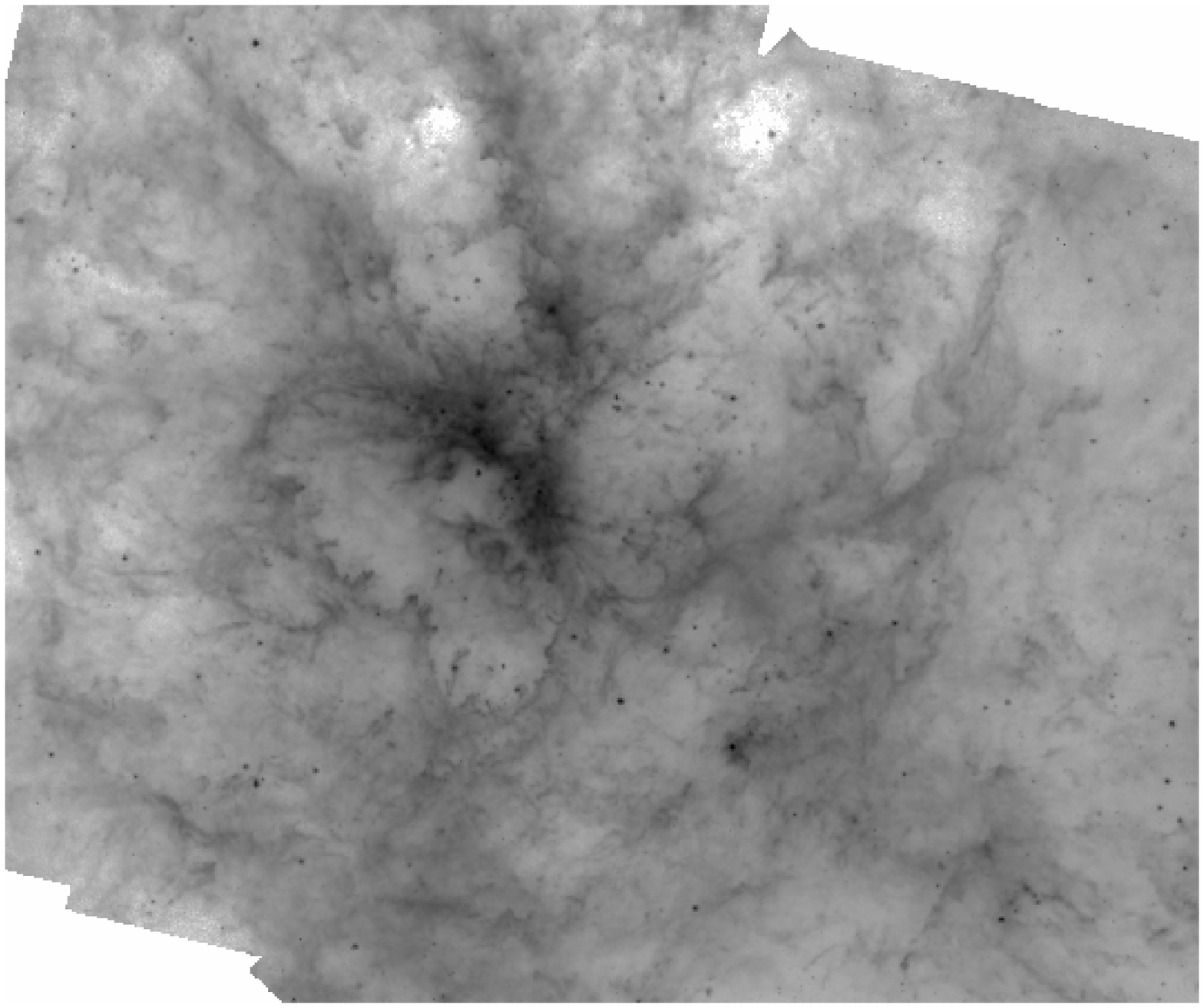,width=6.0cm,angle=0}
\psfig{figure=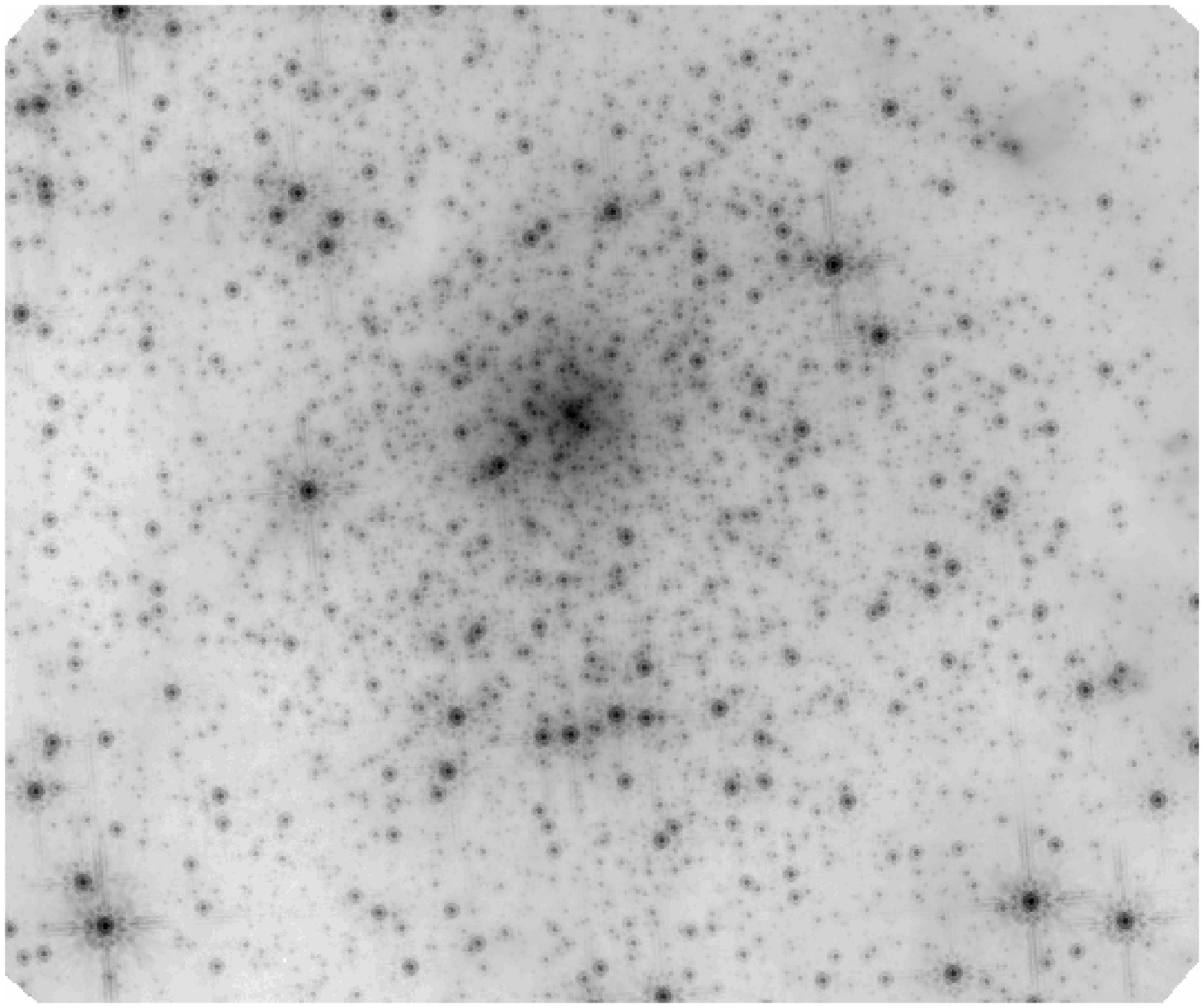,width=6.0cm,angle=0} }} 
\sidebyside 
{\letteredcaption{a}{Spitzer-IRAC image at $8\mu$m showing the
                     distribution of warm dust and UV excited PAH
                     molecules in \30.  The size is $27'\times 20'$
                     (360\,pc~$\times$~270\,pc)
                     \cite{bra05}.\label{spitzer}}}
{\letteredcaption{b}{HST-NICMOS $H$-band image of the $90''\times
                     75''$ (23\,pc~$\times$~19\,pc) around R136
                     \cite{and05}; the faintest stars visible are
                     about $2 M_\odot$.\label{nicmos}}}
\end{figure}

The most relevant properties of its stellar content and instellar
medium (ISM) are summarized in Tab.~\ref{prop}.  Fig.~\ref{spitzer}
shows the complex nature of the ISM in \30 with its interplay between
large-scale filaments and wind-blown cavities.  According to
\cite{wal97} NGC\,2070 consists of several distinct, young stellar
generations.  The IMF is fully populated up to at least $100 M_\odot$
\cite{mas98} and there is no evidence for a truncated low-mass IMF,
even within R136 \cite{and05}.  The stellar distribution near the center
R136 is shown in Fig.~\ref{nicmos}.

\small
\begin{table}[ht]
\caption{Properties of the stars and the ISM in \30 \label{prop}}
\begin{tabular*}{\textwidth}{lcl}
\sphline
\it Quantity &\it Value &\it Reference \cr
\sphline
H$\alpha$ luminosity & $1.5\times 10^{40}$ erg\ s$^{-1}$ & \cite{ken84}\cr
Ly-cont flux (\30) & $1.1\times 10^{52}$ phot\ s$^{-1}$ & \cite{ken84}\cr
Ly-cont flux (NGC\,2070) & $4.5\times 10^{51}$ phot\ s$^{-1}$ & \cite{wal91}\cr
\# OB stars (NGC\,2070) & 2400 & \cite{par93}\cr
\sphline
H$_2$ mass & $7\times 10^{7} M_\odot$ & \cr
\h2 mass   & $8\times 10^{5} M_\odot$ & \cite{ken84} \cr
E$_{\mbox{kin}}^{\mbox{gas}}$ & $\ge 10^{52}$ erg & \cite{chu94} \cr
L$_{\mbox{FIR}}$ & $4\times 10^{7} L_\odot$ & \cite{wer78}\cr
\sphline
\end{tabular*}
\end{table}
\normalsize


\section{Is R136 a Super Star Cluster (SSC)?}

With about 40 stars of spectral type O3 R136 is the densest
concentration of very massive stars known \cite{mas98}.  R136 is
centrally condensed and displays a remarkably smooth exponential
cluster profile (see e.g., \cite{mal94}, their Fig.~13).  The stellar
properties of R136 are summarized in Tab.~\ref{stellar}.  

\small
\begin{table}[ht]
\caption{Stellar properties of R136\label{stellar}}
\begin{tabular*}{\textwidth}{lcl}
\sphline
\it Quantity &\it Value &\it Reference \cr
\sphline
age & $\approx 2\pm 1$ Myr & $^a$ \cr
central density $\rho_c$ & $5.5\times 10^{4} M_\odot$ pc$^{-3}$ & 
    \cite{hun95}\cr
total mass $m_{tot}$ & $6.3\times 10^{4} M_\odot^b$ & 
    \cite{hun95}\cr
core radius $r_c$ & 0.12 pc  ($0.\!''5$) & \cite{bra96}\cr
half-mass radius $r_{hm}$ & 1.2 pc  ($5''$) & \cite{bra96}\cr
tidal radius $r_t$ & 5 pc  ($21''$) & \cite{mey93}\cr
IMF slope $\xi$ & $2.2^c$ & \cite{and05}\cr
\sphline
\end{tabular*}
\begin{tablenotes}
$^a$ the age of the cluster is still subject of controversy

$^b$ for $m \ge 0.1 M_\odot$

$^c$ the \cite{sal55} slope is $\xi = 2.35$ in this notation
\end{tablenotes}
\end{table}
\normalsize

However, there are numerous examples of massive young clusters that do
{\sl not} show a density profile like R136.  For instance, NGC\,604,
the most luminous \h2 region in M33, looks very similar to R136 in H$\alpha$
\cite{hun96} both in structure and luminosity, but the stellar distribution
is completely different, given by numerous smaller
clusters --- a structure sometimes referred to as a scaled
OB~association (SOBA) \cite{mai04}.

\begin{figure}[ht]
\centerline{\hbox{
\psfig{figure=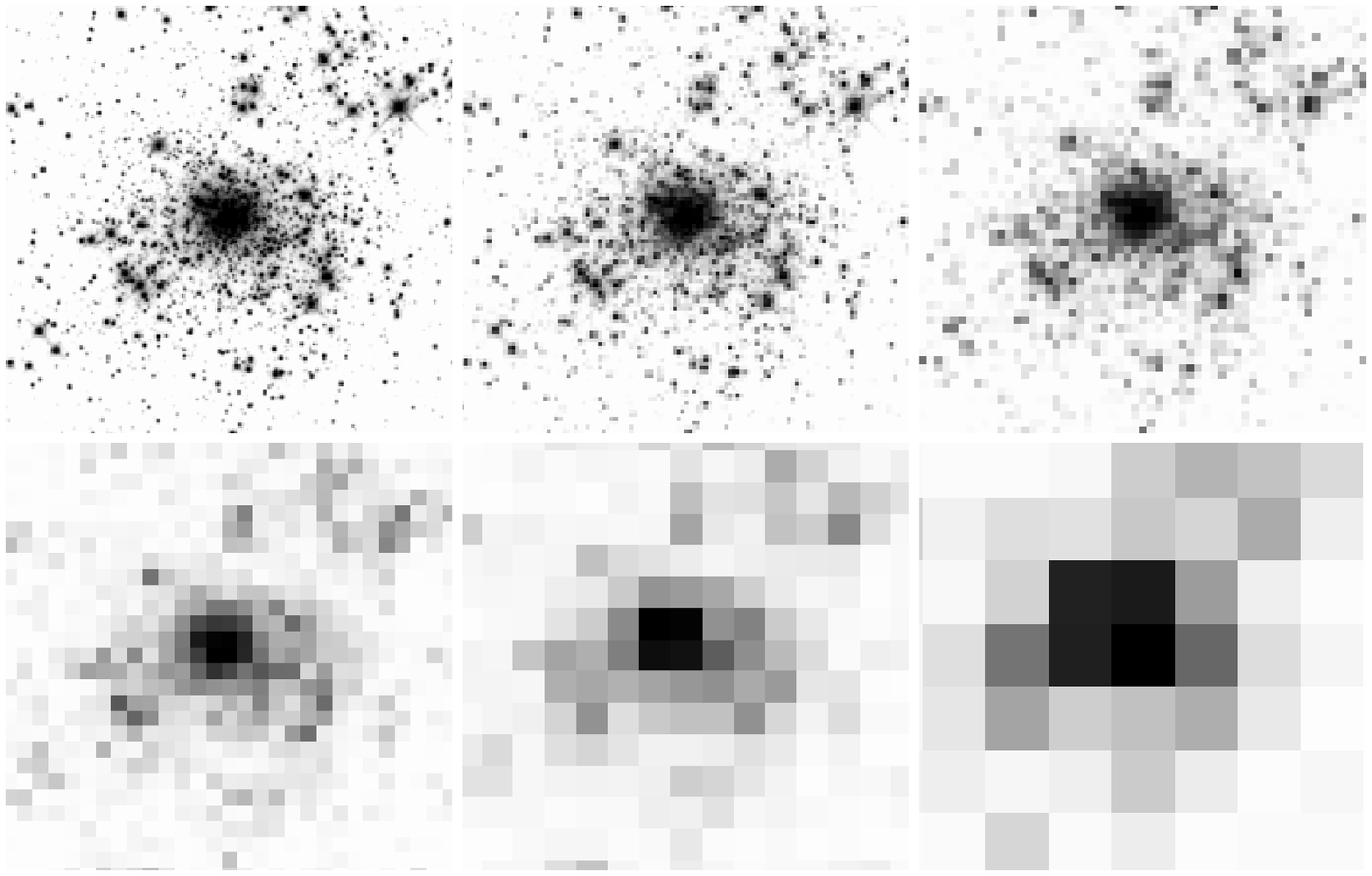,width=6.0cm,angle=0}
\psfig{figure=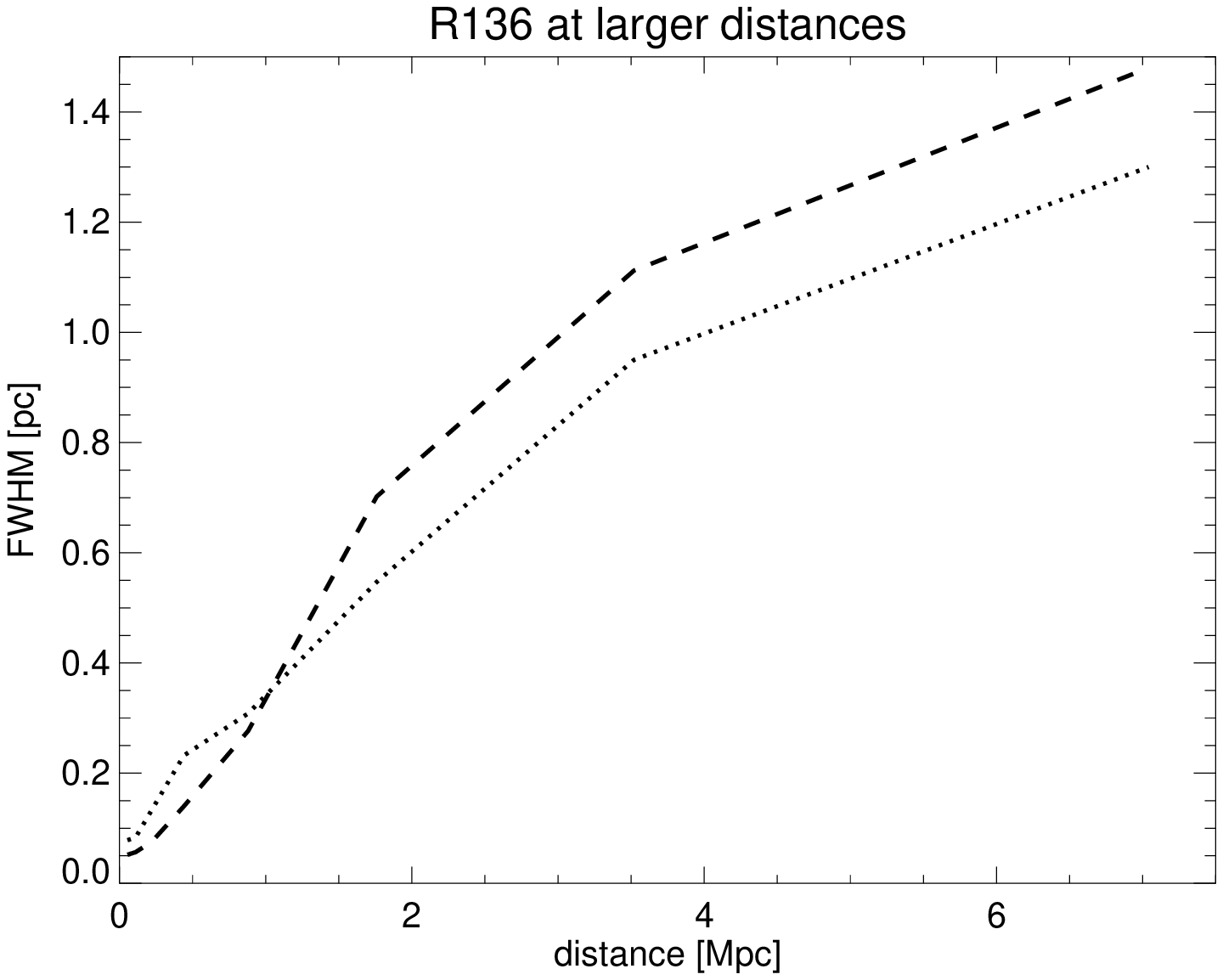,width=6.0cm,angle=0}
}}
\sidebyside
{\letteredcaption{a}{R136 at the observed NICMOS resolution (upper left)
                     and at larger distances that are increasing by 
                     factors of 2, 4, 8, 16, and 32.\label{rebin}}}
{\letteredcaption{b}{The FWHM derived from 2D Gaussian fits to the 
                     light profiles of R136 in Fig.~\ref{rebin}.
                     \label{fwhm}}}
\end{figure}

In recent years, so-called super star clusters, such as in the
Antennae galaxies, have received a lot of attention (e.g.,
\cite{men02}).  The dynamical mass can be estimated via $m_{dyn} =
\eta \frac{\sigma^2 r_{hl}}{G}$, where $\eta \approx 10$.  However,
this is based on the assumption that the cluster is well resolved,
i.e., that $r_{hl}$ can be accurately determined.  Fig.~\ref{rebin}
shows R136 in the upper left as observed with NICMOS (same as
Fig.~\ref{nicmos}), and then progressively at $2\times$ lower
resolution.  The FWHM derived from the same cluster but at different
resolution is plotted as a function of distance in Fig.~\ref{fwhm}.
Because of spatial undersampling and the light from bright stars near
the cluster core, the half-light radius $r_{hl}$ --- and thus
$m_{dyn}$ --- can easily be overestimated by a factor of ten or more
at distances beyond a few Mpc.  To complicate matters, the
spectroscopically measured velocity dispersion may be significantly
affected by the orbital velocities of massive binary stars
\cite{bos01}.  Furthermore, at distances like the Antennae, a
centrally condensed cluster like R136 is indistinguishable from a
non-virialized, NGC\,604-like SOBA.  Given the large uncertainties, it
remains to be seen how much more ``super'' than R136 super star
clusters really are.

\section{Are luminous starbursts made of 30 Doradus complexes?}

\30 does not experience a strong gravitational field from its host
galaxy, the entire complex can almost freely expand, and its center
has no continuous supply from a larger gas reservoir.  Under different
boundary conditions, which may be given in a circum-nuclear starburst,
it is conceivable that the densities of gas, dust, and embedded star
clusters are much higher --- without necessarily requiring different
unit cells than R136.  Using a very simplified model of identical
clusters, spherical geometry, and constant gas density, we can
estimate what the total luminosity of such a starburst might be.

Assuming 200~pc for the linear extend of a ``typical'' extragalactic
starburst region, and taking R136's tidal radius $r_t$ as the
(half-)size of a unit cell, there could be the equivalent of as many
as 8000 R136-like clusters within the given volume.  If only a quarter
($10^7 M_\odot$) of the total \30 far-IR luminosity is being produced
within the volume defined by $r_t$, and if $L_{FIR}$ is the sum of the
reprocessed UV-radiation and (about the same amount) of shock-heated
gas from supernovae, the total far-infrared luminosity is:
\begin{equation}
L_{FIR}^{tot} = 8000 \times 2 \times 10^7 = 1.6\times 10^{11} L_\odot
\end{equation}

This number is well within the ballpark of luminous starburst
galaxies, although it falls short of the luminosity of ultra-luminous
infrared galaxies.  At any rate, if starbursts have such porous,
inhomogenous structures they represent a big challenge for accurate
starburst modelling.



\begin{chapthebibliography}{1}
\bibitem[(Andersen et al. 2005)]{and05} Andersen, M.,
         Brandl, B.R. \& Zinnecker, H. 2005, ApJ, in preparation
\bibitem[(Bosch et al. 2001)]{bos01} Bosch, G. et al. 2001, A\,\&\,A, 380, 137
\bibitem[Brandl et al. (1996)]{bra96} Brandl, B.R. et al. 1996, ApJ, 466,
         254
\bibitem[(Brandl et al. 2005)]{bra05} Brandl, B.R. et al. 2005, ApJ,
         in preparation
\bibitem[Chu \& Kennicutt (1994)]{chu94} Chu, Y.-H. \& Kennicutt,
         R.C. 1994, ApJ, 425, 720
\bibitem[Hunter et al. (1995)]{hun95}
         Hunter, D. et al. 1995, ApJ, 448, 179
\bibitem[(Hunter et al. 1996)]{hun96}
         Hunter, D. et al. 1996, ApJ, 456, 174
\bibitem[Kennicutt (1984)]{ken84}
         Kennicutt, R. C.  1984, ApJ, 287, 116
\bibitem[(Ma\'{i}z-Apell\'{a}niz et al. 2004)]{mai04} 	
 	 Ma\'{i}z-Apell\'{a}niz, J. et al. 2004, AJ, 128, 1196
\bibitem[Malumuth \& Heap (1994)]{mal94}
         Malumuth, E.M. \& Heap, S.R. 1994, AJ, 107, 1054
\bibitem[(Massey \& Hunter 1998)]{mas98}
         Massey \& Hunter 1998, ApJ, 493, 180
\bibitem[Mengel et al. (2002)]{men02} Mengel, S. et al. 2002, A\,\&\,A, 383,
         137
\bibitem[Meylan (1993)]{mey93}
         Meylan 1993, ASP, vol. 48, 588
\bibitem[Moffat (1994)]{mof94}
         Moffat 1994, ApJ, 436, 183
\bibitem[Parker (1993)]{par93}
         Parker, J.W. 1993, AJ, 106, 560
\bibitem[Salpeter (1955)]{sal55} Salpeter, E.E. 1955, ApJ, 123, 666
\bibitem[Walborn (1991)]{wal91} Walborn, N.R. 1991, IAUS 148, 145
\bibitem[Walborn \& Blades (1997)]{wal97} Walborn, N.R. \& Blades,
         J.C. 1997, ApJS, 112, 457
\bibitem[Werner et al. (1978)]{wer78} Werner, M.W. et al. 1978, MNRAS, 184, 365

\end{chapthebibliography}
\end{document}